\documentstyle[12pt]{article}

\newcommand{\Od}{{\cal O}}

\def\lsim{\raisebox{-.4ex}{$\stackrel{<}{\scriptstyle \sim}$\,}}
\def\gsim{\raisebox{-.4ex}{$\stackrel{>}{\scriptstyle \sim}$\,}}

\begin{document}
%


\input epsf \renewcommand{\topfraction}{0.8}
\pagestyle{empty} \vspace*{-10mm}

\begin{flushright}
\normalsize{UCI-TR-2005-37}
\end{flushright}

\begin{center}
\Large{\bf Branon radiative corrections to collider physics and precision observables}
\\ \vspace*{2cm}

\large{ J. A. R. Cembranos$^{1}$, A. Dobado$^{2}$ and A. L.
Maroto$^2$}
\\ \vspace{0.2cm} \normalsize
$^1$ Department of Physics and Astronomy,\\
 University of California, Irvine, CA 92697, USA\\
$^2$ Departamento de  F\'{\i}sica Te\'orica,\\
 Universidad Complutense de
  Madrid, 28040 Madrid, Spain\\
\vspace*{0.6cm} {\bf ABSTRACT} \\
\end{center} \vspace*{5mm}

In the context of brane-world scenarios, we study the  effects
produced by the exchange of virtual massive branons. A one-loop
calculation is performed which  generates higher-dimensional
operators involving SM fields suppressed by powers of the brane
tension scale. We discuss constraints on this scenario from
colliders such as HERA, LEP and Tevatron and prospects for future
detections at LHC or ILC. The most interesting
phenomenology comes from new four-particles vertices induced by
branon radiative corrections, mainly from four fermion
interactions. The presence of flexible branes modifies also the
muon anomalous magnetic moment and the electroweak precision
observables.

 \noindent
\begin{flushleft} PACS: 11.25Mj, 11.10Lm, 11.15Ex \\
\end{flushleft}
\newpage
\setcounter{page}{1} \pagestyle{plain}
\textheight 20 true cm
\section{Introduction}

In recent years, it has been shown that a generic property of
brane-world models \cite{ADD} with low tension ($\tau \equiv f^4
\ll \Lambda^4$, where $\tau$ is the brane tension and $\Lambda$ is
the scale below which the description given by the brane-world
scenario is appropriate) is the presence of new modes
$\pi^\alpha(x)$ called branons which roughly correspond to
excitations of the brane position along the extra compactified
dimensions. The relevant tree-level phenomenology  of
branons has been studied for colliders and also for astrophysics
and cosmology in terms of their mass $M$ and brane tension
parameter $f$, and it has been suggested that massive branons
could be natural candidates for dark matter in this kind of models
\cite{CDM}.

In this work we study the phenomenology of branon radiative
corrections. Branon loops are interesting mainly for two reasons.
First because precision tests of the Standard Model (SM) usually
enforce strong constraints on physics beyond it and thus make
possible to reject many new models, or at least to set bounds on
their parameters. The second reason is that branon loops provide
new physical effects, such as four fermion interactions, which can be
searched for in present and next generation colliders.

As it is the case for branon tree-level effects, the loop
corrections can be obtained from the effective action for branons
described in detail in \cite{DoMa}. This effective action can be
expanded in  powers of $\partial \pi/f^2$ and
$M^2\pi^2/f^4$ \cite{ACDM,BSky}:
\begin{equation}
S_{eff}[\pi]=S_{eff}^{(0)}[\pi]+ S_{eff}^{(2)}[\pi]+
S_{eff}^{(4)}[\pi]+ ...
\end{equation}
The zeroth order term is just a constant, the $\Od((\pi/f)^2)$
contribution contains the branon free action:
\begin{eqnarray}
 S_{eff}^{(2)}[\pi]=\frac{1}{2}\int_{M_4}d^4x
(\delta_{\alpha\beta}\partial_{\mu}\pi^\alpha\partial^{\mu}\pi^\beta
-M^2_{\alpha\beta}\pi^\alpha\pi^\beta).
\end{eqnarray}
where $M^2_{\alpha\beta}$ is the squared branon mass matrix corresponding
to the different branon excitations $\pi^\alpha$, with $\alpha$ running
from one
to the number of effective extra dimensions $N$. The couplings to
the SM fields $\Phi$ living on the brane (or any suitable
extension of it) in the presence of a gravitational background
which for simplicity we will assume to be flat
($g_{\mu\nu}=\eta_{\mu\nu}$), can be described at low energies by
the action:
\begin{eqnarray}
S_{SM}[\Phi,\pi] &=&\int_{M_4}d^4x\left[ {\mathcal L}_{SM}(\Phi) +
\frac{1}{2}\delta_{\alpha\beta}\partial_{\mu}\pi^\alpha
\partial^{\mu}\pi^\beta-\frac{1}{2}M^2_{\alpha\beta}\pi^\alpha\pi^\beta
\right.
\nonumber\\
&+& \left.\frac{1}{8f^4}(4\delta_{\alpha\beta}\partial_{\mu}\pi^\alpha
\partial_{\nu}\pi^\beta-M^2_{\alpha\beta}\pi^\alpha\pi^\beta\eta_{\mu\nu})
T^{\mu\nu}_{SM} \right]
+{\cal O}(\pi^4).
\label{quadratic}
\end{eqnarray}
where $T^{\mu\nu}_{SM}$ is the conserved SM
energy-momentum  evaluated in the background metric:

\begin{eqnarray}
T^{\mu\nu}_{SM}=-\left.\left( g^{\mu\nu}{\cal
L}_{SM}+2\frac{\delta {\cal L}_{SM}}{\delta
g_{\mu\nu}}\right)\right \vert_{ g_{\mu\nu}=\eta_{\mu\nu}}
\end{eqnarray}

It is interesting to note that there is no single branon
interactions due to the parity conservation on the brane by the
 gravitational action. Thus branons are absolutely stable.
This fact is crucial for the branon phenomenology
and in particular for cosmology  since it makes them natural
WIMP candidates for dark matter \cite{CDM}. In addition, the quadratic
expression in (\ref{quadratic}) is valid for any internal extra-dimension
space $K_N$, regardless  the particular form of its metric 
$\gamma_{\alpha\beta}$.
Indeed the form of the couplings only depends on the number $N$ of  branon
fields and the brane tension. Dependence on the geometry of the extra
dimensions will appear only at higher orders. Here we are assuming that the
bulk $D$ dimensional space-time ($D=4+N$) can be split as
${\cal M}_D=M_4 \times K_N$, where $M_4$ is the standard four-dimensional
Minkowski space and $K_N$ is some compact and homogeneous space of dimension
$N$ with gaussian coordinates $y^\alpha$. Then the $N$ branon fields
($\alpha=1,...,N$) can
be chosen so that $\pi^\alpha (x)= f^2 y^\alpha(x)$
where $y^\alpha=y^\alpha(x)$ represents
the position  of the excited brane in the extra-dimension space $K_N$. The
brane ground state corresponds to $\pi^\alpha=0$.

From the action above it is clear that branons always interact by
pairs with the SM matter fields. In addition, due to their
geometric origin, those interactions are very similar to the
gravitational ones since the $\pi^\alpha$ fields couple to all the matter
fields through the energy-momentum tensor and with the same
strength suppressed by a $f^4$ factor. The interaction between
 bulk gravitons and SM fields is given by:
\begin{eqnarray}
S_h&=& \frac{1}{\bar M_P} \sum_p  \int_{M_4}d^4x
h_{\mu\nu}^{(p)}(x) T^{\mu\nu}_{SM}(x)
\end{eqnarray}
where $\bar M_P^2 \equiv M_P^ 2/ 4 \pi$ is the squared reduced
Planck mass $(M_P=1.2 \times 10^{19}$ GeV) and $h_{\mu\nu}^{(p)}$
are the Kaluza-Klein (KK) modes of the bulk graviton
$h_{\mu\nu}(x,y)$ corresponding to the $4\times 4$ part of the
bulk metric:
\begin{equation}
g_{\mu\nu}(x,y)=\eta_{\mu\nu}+\frac{2h_{\mu\nu}(x,y)}{\bar
M_D^{1+N/2}}
\end{equation}
where $\bar M_D^{D-2}= M_D^{D-2}/(4 \pi)$ and  
$M_D$ is the $D$ dimensional Planck scale (fundamental scale
of gravity) related with the usual four-dimensional
Planck scale by $M_P^2=V(K_N)M_D^
{D-2}$ with $V(K_N)$ being the volume of the internal space $K_N$
(notice that $M_4\equiv M_P$). By introducing a complete orthonormal
set of functions $f_p=f_p(y)$ on $K_N$ with normalization:
\begin{equation}
\int _{K_n} d\mbox{V}(K_N)\,f_p^*(y)f_q(y)= V(K_n)\delta_{pq}
\end{equation}
and $f_p(0)=1$, the KK mode decomposition for the graviton field becomes
\begin{equation}
h_{\mu\nu}(x,y)=\frac{1}{\sqrt{V(K_N)}}\sum_p h_{\mu\nu}^
{(p)}(x)f_p(y).
\end{equation}

When computing radiative corrections, divergent integrals appear. As
our effective actions are not renormalizable all our results will
be given in terms of some energy cut-off $\Lambda$, which could be
taken as the value where the whole brane-world picture breaks down
and a more fundamental approach is needed. Then our results will
be given in terms of four parameters, namely the number of branons
or extra dimensions $N$, the  branon mass $M$ (for simplicity we
will assume at the end that all of them are degenerate
$M_{\alpha\beta}=\delta_{\alpha\beta}M_\beta=\delta_{\alpha\beta}M$), the brane tension scale $f$
($\tau=f^ 4$) and the cut-off $\Lambda$.

The plan of the paper goes as follows: In section 2 we reobtain
the result of \cite{GB} concerning  the suppression of the
coupling between SM fields on the brane and bulk fields, by
integrating out the branon fields instead of using
arguments based on normal ordering. In section 3 we study the
effects of branon loops on the SM particle parameters and find the
effective action describing the new induced interactions. The
corresponding phenomenological consequences are considered in
section 4, where we also set the bounds coming  from the
branon loops on the parameters $f$, $M$, $N$ and the scale
$\Lambda$. Further constraints can be obtained from two loops
effects and their impact on the electroweak precision observables
and the muon anomalous magnetic momentum which can be found in
section 5. In section 6 we summarize and comment our results and in
Appendix A, B and C we define the divergent integrals appearing in our
computations, the Feynman rules corresponding to the effective
Lagrangian describing the branon loops effects, and the associated
cross-sections.

\section{Graviton coupling suppression}

Probably the most immediate effect of virtual branons is the
suppression of the coupling of SM particles and the KK modes bulk
fields like the graviton. When branon fluctuations are taken into
account this effective coupling is described by the action:
\begin{eqnarray}
S_h&=& \frac{1}{\bar M_P} \sum_p  \int_{M_4}d^4x
h_{\mu\nu}^{(p)}(x) T^{\mu\nu}_{SM}(x)f_p(\pi)
\end{eqnarray}
due to the fact that the brane is no more sitting at $\pi=0$ but
is moving around this point. Now the branons fields can be
integrated out in the usual way to find:
\begin{eqnarray}
S_h&=& \frac{1}{\bar M_P} \sum_p  \int_{M_4}d^4x
h_{\mu\nu}^{(p)}(x) T^{\mu\nu}_{SM}(x)\langle f_p(\pi)\rangle
\end{eqnarray}
where the $f_p$ expectation value is given by
\begin{eqnarray}
\langle f_p(\pi) \rangle = \int [d\pi] e^{i S_{eff}^{(2)}[\pi]} f_p(\pi)
\end{eqnarray}
In the limit of massless branons, the branon effective action is
just a non linear sigma model (NLSM) based on a coset which is
isomorphic to $K_N$. Therefore the invariant path integral measure
should include an additional factor proportional to the square
root determinant of the coset metric to ensure that quantum
corrections do not spoil the Ward identities of the NLSM. The
extra term in the measure amounts to an extra term in the
effective action proportional to $\Lambda^4$. This term is
important when dealing with branon loop corrections to the branon
self-interactions (for instance branon-branon elastic scattering)
\cite{Espriu}. However in this work we are mainly interested in interactions
between a couple of branons and SM particles and hence we can
safely neglect this measure term.

To compute the path integral above, we need to know the precise form
of the $f(\pi)$ functions which depends on the $K_N$ geometry. For
example for the case of the torus $K_N= T^N$:
\begin{equation}
f_{\vec n}(y)=\exp \left(i\frac{\vec n \vec y}{R}\right)
\end{equation}
where $\vec n=(n_1,n_2,...n_N)$ is a $N$ dimensional vector with integer and
positive or zero components and $R$ is the torus radius (common
for all coordinates). Then we have
\begin{equation}
\langle f_{\vec n}(\pi)\rangle = \langle
\exp \left(i\frac{\vec n \vec \pi}{R f^ 2}\right)\rangle =\exp
\left(-\frac{1}{2R^ 2 f^ 4}\sum_{\alpha=1}^N n_\alpha^ 2
G_{\alpha\alpha}(0)\right)
\end{equation}
where  $G_{\alpha\beta}(x)$ is the branon propagator
\begin{eqnarray}
G_{\alpha\beta}(x-y)=\int d\tilde q
e^{-iq(x-y)}\frac{\delta_{\alpha\beta}}{q^2-M_\alpha^2+i\epsilon}
\end{eqnarray}
and  $d\tilde q\equiv d^4q/(2\pi)^4$. Using a cut-off $\Lambda$ to
regularize the divergent integral we find
\begin{equation}
G_{\alpha\beta}(0)=\frac{1}{16\pi^ 2}
\left[\Lambda^ 2-M_\alpha^ 2\log \left(\frac{\Lambda^
2}{M_\alpha^ 2}+1\right)\right]\delta_{\alpha\beta}
\end{equation}
Then the effective action becomes
\begin{eqnarray}
S_h&=& \frac{1}{\bar M_P} \sum_{\vec n}  \int_{M_4}d^4x g_{\vec
n}h_{\mu\nu}^{(\vec n)}(x) T^{\mu\nu}_{SM}(x)
\end{eqnarray}
In other words, the effect of branon quantum fluctuations amounts
to introducing the KK mode dependent couplings $g_{\vec n}$ which
are given for toroidal compactification by
\begin{equation}
g_{\vec n}=\exp  \left(-\frac{\Lambda^ 2}{32\pi^2R^2 f^4}
\sum_{\alpha=1}^N
n_\alpha^ 2\left[1-\frac{M_\alpha^ 2}{\Lambda^ 2}
\log \left(\frac{\Lambda^ 2}{M_\alpha^
2}+1\right)\right]\right)
\end{equation}
Thus the coupling of SM matter to higher KK modes is exponentially
suppressed.  This result was first obtained in \cite{GB} for
massless branons by using an argument based on normal ordering.
Our derivation here is more natural in the context of the path
integral treatment of branon quantum fluctuations, and also it can
be applied to massive branons in any extra dimension space $K_N$.
In any case this coupling suppression  has very interesting
consequences from the phenomenological point of view. It improves
the unitarity behavior of the cross section for producing
gravitons from SM particles and, in addition, it solves the problem
of the divergences appearing even at the tree level when one
considers the KK  graviton tower propagators for dimension equal or
larger than two. Moreover whenever we have $v\equiv R f^ 2 \ll
\Lambda$, KK gravitons decouple from the SM particles, so that at
low energies the only brane-world related particles that must be
taken into account are branons. In the following we will assume
this to be the case and accordingly we will deal only with SM particles
and branons.

\section{Branon-loops effects on SM particles}

In order to study the effect of virtual branons on the SM
particles, it is useful to introduce the SM effective action
$\Gamma_{SM}^{eff}[\Phi]$ obtained after integrating out the
branon fields:
\begin{eqnarray}
e^{i\Gamma_{SM}^{eff}[\Phi]}=\int [d\pi] e^{iS_{SM}[\Phi]}
=e^{i\int d^4x {\cal L}_{SM}}(\mbox{Det}[O])^{-1/2}
\end{eqnarray}
where the $O$ operator is defined as $O=A+B$ with:
\begin{eqnarray}
A_{\alpha\beta}(x,y)=-\delta_{\alpha\beta}[
\partial_\mu\partial^\mu+M_\alpha^2]\delta(x-y).
\end{eqnarray}
\begin{eqnarray}
B_{\alpha\beta}(x-y)=\frac{-1}{f^4}\delta_{\alpha\beta}\left[
T_{SM}^{\mu\nu}(\partial_\mu\partial_\nu+\frac{M_\alpha^2}{4}
\eta_{\mu\nu})\right]
\delta(x-y),
\end{eqnarray}
This SM effective action can be computed in a systematic way by
using standard procedures (see for instance \cite{Book}). Thus we
have:
\begin{eqnarray}
\Gamma^{eff}_{SM}[\Phi]=\int d^4x {\cal
L}_{SM}+\frac{i}{2}\rm{Tr}(\ln[O]),
\end{eqnarray}
and
\begin{eqnarray}
\rm{Tr}(\ln[O])=\rm{Tr}(\ln[A])+\rm{Tr}(\ln[ 1+B A^{-1}]).
\end{eqnarray}
The first term does not depend on the SM fields and it can only
contribute  to  the renormalization of the
  the cosmological constant. Expanding the logarithm, we obtain the usual expression:
\begin{eqnarray}
\Gamma^{eff}_{SM}[\Phi]&=& \int d^4x {\cal L}_{SM}
-\frac{i}{2}\sum_{k=1}^\infty\frac{(-1)^k}{k} \rm{Tr}(B A^{-1})^k\nonumber\\
&=&\int d^4x {\cal L}_{SM}+\sum_{k=1}^\infty\Gamma^{(k)}[\Phi],
\end{eqnarray}
where $A^{-1}$ is the branon ($\pi^\alpha$) free propagator:
\begin{eqnarray}
A_{\alpha\beta}^{-1}(x,y)= G_{\alpha\beta}(x-y).
\end{eqnarray}
Then the first contribution $\Gamma^{(1)}$, reads:
\begin{eqnarray}
\Gamma^{(1)}[\Phi]&=&\frac{i}{2}\int d^4x\, d^4y\,
B^{\alpha\beta}(x,y)G_{\alpha\beta}(x-y)
=C_1\int d^ 4x\, T^{\,\,\mu}_{SM\,\,\mu},
\end{eqnarray}
%
where, assuming all the branons to be degenerate
($M_\alpha=M,\; \alpha=1,\dots, N$), the $C_1$
constant is given by:
%
\begin{eqnarray}
C_1\eta_{\mu\nu}&=& i\frac{N}{8f^4}\int d\tilde q \frac{4q_\mu
q_\nu - M^2\eta_{\mu\nu}}{q^2-M^2+i\epsilon}.
\end{eqnarray}
The second contribution to the effective action is:
\begin{eqnarray}
\Gamma^{(2)}[\Phi]&=&-\frac{i}{4}\int d^4x\, d^4y\, d^4z\,d^4t \,
B_{\alpha\beta}(x,y)G_{\beta\gamma}(y-z)B_{\gamma\delta}(z,t)
G_{\delta\alpha}(t-x)
\nonumber\\
&=&-i\frac{N}{4f^8}\int d^ 4x\, d^ 4y\, d^ 4p e^{-ip(y-x)}
T^{\mu\nu}(x)T^{\rho\sigma}(y)\\
& &\left[J^{(M)}_{\mu\nu\rho\sigma}(p)
-\frac{M^2}{4}(\eta_{\mu\nu}J^{(2)}_{\rho\sigma}(p)+
\eta_{\rho\sigma}J^{(2)}_{\mu\nu}(p))
+\frac{M^4}{16}\eta_{\mu\nu}\eta_{\rho\sigma}J^{(0)}(p)\right]\nonumber
\end{eqnarray}
where from now on $T^{\mu\nu}=T^{\mu\nu}_{SM}$ and the integrals
$J^{(I)}$ are defined in  Appendix \ref{Integrals}. It is
 convenient to split  the final expression into a local divergent
  term $\Gamma^{(2)}_{L}[\Phi]$
  and a non-local finite term: $\Gamma^{(2)}_{NL}[\Phi]$.
\begin{eqnarray}
\Gamma^{(2)}[\Phi]&=&\Gamma^{(2)}_{L}[\Phi]
+\Gamma^{(2)}_{NL}[\Phi]. \label{result}
\end{eqnarray}
Now by using  the equation of motion at the zero order:
$\partial_\mu T_{SM}^{\mu\nu}=0$, the local piece can be written
in terms of six constants $W_i$, ($i=1,2,.. 6$):
\begin{eqnarray}
\Gamma^{(2)}_{L}[\Phi]&=&\int dx\,
\{W_1 T^{\mu\nu}T_{\mu\nu}+W_2 T^{\mu}_\mu T^\nu_{\nu}
\nonumber\\
&+&W_3 T^{\mu\nu}\Box T_{\mu\nu}+
W_4 T^{\mu}_\mu\Box T^\nu_{\nu}
\nonumber\\
&+&W_5 T^{\mu\nu}\Box^2 T_{\mu\nu}+
W_6 T^{\mu}_\mu\Box^2 T^\nu_{\nu}\}
\label{local}
\end{eqnarray}
and the non-local one in terms of two functions $D_j(p)$,
($j=1,2$):
\begin{eqnarray}
\Gamma^{(2)}_{NL}[\Phi]&=&\int dx\, dy\, dp e^{-ip(y-x)}
\nonumber\\
&&\{D_1(p)T^{\mu\nu}(x)T_{\mu\nu}(y)
+D_2(p)T^{\mu}_\mu(x)T^\nu_{\nu}(y)\}.
\label{nonlocal}
\end{eqnarray}
These two functions are given by:
\begin{eqnarray}
D_1(p)&=&\frac{-iN}{480f^8}\,
      \,{\left(p^2-4\,M^2\right)}^2\,J^F(p,M),
\label{nonlocal1}
\end{eqnarray}
\begin{eqnarray}
D_2(p)&=&-i\frac{N}{960f^8}\,\{
      \left(p^2+6\,M^2\right)\left(p^2-4\,M^2\right)
+15M^4\}J^F(p,M),
\label{nonlocal2}
\end{eqnarray}
with
\begin{eqnarray}
\label{jf}
 J^F(p,M)=\left\{
\begin{array}{ccc}
\frac{i}{(4\pi)^2}\left[2-\sqrt{1-\frac{4M^2}{p^2}}
\ln\left(\frac{\sqrt{1-\frac{4M^2}{p^2}}+1}
{\sqrt{1-\frac{4M^2}{p^2}}-1}\right)\right]
&;&p^2\leq 0\, , \\
\frac{i}{(4\pi)^2}\left[2+2\sqrt{\frac{4M^2}{p^2}-1}
\tan^{-1}\left(\frac{1}{\sqrt{\frac{4M^2}{p^2}-1}}\right)\right]
&;&0<p^2\leq 4M^2\, , \\
\frac{i}{(4\pi)^2}\left[2-\sqrt{1-\frac{4M^2}{p^2}}
\ln\left(\frac{1+\sqrt{1-\frac{4M^2}{p^2}}}
{1-\sqrt{1-\frac{4M^2}{p^2}}}\right)+i\pi\right]
&;&4M^2<p^2.
\end{array}\right.
\nonumber\\
\end{eqnarray}

Thus for the particular case of massless branons we find:

\begin{eqnarray}
D_1(p)=2D_2(p)&=&\frac{-iNp^4 J^{F}(p,M=0)}{480f^8}
=-\frac{Np^4\ln(-p^2)}{480(4\pi)^2f^8}
\end{eqnarray}
which is in agreement with previous results \cite{Kugo, CrSt}.

\begin{table}
\centering
\begin{tabular}{||c|cc||}
\hline\hline
Coefficient &Cut-off regularized value &\\
\hline
$W_1$ & $\frac{N\,\left( \Lambda^4 - 4\,\Lambda^2\,M^2 - 2\,M^4 -
6\,M^4\,\ln (\frac{M^2}{\Lambda^2}) \right) }{96\,(4\pi)^2\,f^8}$&
\\
$W_2$ & $\frac{N\,\left( {\left( \Lambda^2 + M^2 \right) }^2 +
3\,M^4\,\ln (\frac{M^2}{\Lambda^2}) \right) }{192\,(4\pi)^2\,f^8}$&
\\
$W_3$ & $\frac{-N\,\left( 15\,\Lambda^2 + 2\,M^2 + 30\,M^2\,\ln
(\frac{M^2}{\Lambda^2}) \right)}{1440\,(4\pi)^2\,f^8}$ &
\\
$W_4$ & $\frac{-N\,\left( 5\,\Lambda^2 - M^2
\right)}{960\,(4\pi)^2\,f^8}$ &
\\
$W_5$ & $\frac{N\,\left( 17 - 60\,\ln (\frac{M^2}{\Lambda^2})
\right)}{28800\,(4\pi)^2\,f^8}$ &
\\
$W_6$ & $\frac{N\,\left( 17 - 60\,\ln (\frac{M^2}{\Lambda^2})
\right)}{57600\,(4\pi)^2\,f^8}$ &
\\
\hline\hline
\end{tabular}
\caption{\label{diver} {Regularized constants computed by using  a
 cut-off $\Lambda$. }}
\end{table}

In the general case,  the local actions $\Gamma^{(1)}[\phi]$ and
$\Gamma^{(2)}_{L}[\phi]$ are divergent and therefore need to be
regularized. By using a cut-off $\Lambda$, the $C_1$  constant
appearing in $\Gamma^{(1)}[\phi]$ is given by:

\begin{eqnarray}
C_1(\Lambda,f)=-\frac{N\Lambda^4}{16(4\pi)^2f^4}.\label{strong}
\end{eqnarray}
The $W_i$ constants can be found in Table 1.
$\Lambda$ could represent the width of brane or any other mechanism
that modified the short-distance theory to cure the ultraviolet
behavior of branons. However, for our purposes, $\Lambda$ is just
a phenomenological parameter. From the point of view of the
effective theory, $\Lambda/f$ parameterizes how strongly (or weakly) coupled quantum
brane is, and therefore controls the unknown relative importance
of tree-level versus loop branon effects. From (\ref{strong}) we
can see that the perturbative loop analysis only makes sense for
approximately $\Lambda\,\lsim\,4\sqrt{\pi}fN^{-1/4}$.

The first term of the effective action $\Gamma^{(1)}[\phi]$ is
 proportional to the trace of the energy momentum tensor so it would
 vanish if the SM were a scale invariant theory. By using the equation
 of motion $\partial_\mu T_{SM}^{\mu\nu}=0$ it is possible to show
 that the only effect of this term is to renormalize the boson (scalars
 or gauge fields)
 masses $m_{\Phi,A}^r=(1-2C_1)^{1/2}m_{\Phi,A}$ and the fermion masses as
$m_\psi^r=(1-C_1)m_\psi$. Therefore, this first correction to the SM
 action does not have any
measurable effect. On the other hand, the local action
$\Gamma^{(2)}_{L}[\phi]$ is important from the phenomenological point of view.
At low enough energies the dominant terms are the ones proportional
to $W_1$ and $W_2$.
Thus the SM Lagrangian is complemented by the additional effective
Lagrangian given by:
\begin{eqnarray}
{\mathcal \triangle L}_{eff}= W_1 T_{\mu\nu}T^{\mu\nu}+W_2 T_\mu^\mu
T_\nu^\nu  \label{eff}.
\end{eqnarray}
where for $\Lambda \gg M$
\begin{eqnarray}
W_1 &=& \frac{N \Lambda^4}{96(4\pi)^2f^8}
\nonumber  \\
W_2 &=& \frac{N \Lambda^4}{192(4\pi)^2f^8}\label{lag}.
\end{eqnarray}
When this is not the case, one should use the full result in Table
1. From this Lagrangian it is possible to obtain the corresponding
Feynman rules (see Appendix \ref{Vertex}). The most relevant
contributions of  branon loops to the SM particle phenomenology
are the four-fermion interactions and fermion pair annihilation
 into two gauge bosons, whose cross-sections can be found in Appendix \ref{cs}.
 
 \section{Phenomenological consequences: constraints}

An effective Lagrangian similar to (\ref{eff})  was
obtained in \cite{CrSt,GS} by integrating at the tree level the
Kaluza-Klein modes of gravitons propagating in the bulk and some
of its phenomenological consequences were studied there. Thus it
is easy to translate some of the results from these references to
the present context.

\begin{table}[bt]
\centering
\begin{tabular}{||c|c c c||} \hline
\hline
Experiment          & $\sqrt s$ (TeV) & ${\cal L}$ (pb$^{-1}$) & $f^2/(N^{1/4}\Lambda)$ (GeV) \\ \hline
HERA$^{\,c}$        & 0.3             &  117                   & 52                           \\
Tevatron-I$^{\,a,\,b}$   & 1.8        &  127                   & 69                           \\
LEP-II$^{\,a}$      & 0.2             &  700                   & 59                           \\
LEP-II$^{\,b}$      & 0.2             &  700                   & 75                           \\ \hline
\multicolumn{3}{||c|}{combined}&81\\
\hline\hline
\end{tabular}
\caption{\label{constrains1} 
{Lower limits on
$f^2/(N^{1/4}\Lambda)$ (in GeV) from 
virtual branon searches at colliders (results 
at the $95\;\%$ c.l.):
HERA \cite{Adloff:2003jm}, LEP-II \cite{unknown:2004qh} and 
Tevatron-I \cite{d0}.
 The indices $^{a,b,c}$ denote the two-photon,
$e^+e^-$ and $e^+p$ ($e^-p$) channels respectively. 
$\sqrt{s}$ is the center of mass energy of the total
process, and ${\mathcal L}$ is the total integrated luminosity.}}
\end{table}


Concerning the four fermion interactions 
$\bar\psi_a(p_1)\psi_a(p_2)
\longrightarrow\bar\psi_b(p_3)\psi_b(p_4)$ (see Apendix B.1.),  
the most interesting case is the Bhabha scattering  at LEP. From the
Lagrangian in (\ref{eff}), it is
possible to find a four-fermion interaction, whose amplitude is given by:
\begin{eqnarray}
{\cal M}^{4\psi}&=& \bar v_{a'}(p_1)\bar u_b(p_4) 
V^{4\psi}_{a'abb'}(-p_1,-p_2, p_3, p_4) 
v_{b'}(p_3)u_a(p_2).
\label{M4f}
\end{eqnarray}
Neglecting  the fermion masses and
assuming that all of them are different, the amplitude is just:
\begin{eqnarray}
{\cal M}^{4\psi}(p_1, p_2, p_3, p_4)
&=&\frac{W_1}{4}\bar v_{a'}(p_1)\bar u_b(p_4)
\biggl[\gamma_{a'a\,\mu}\gamma_{bb'}^\mu(p_2-p_1)_\nu(p_4-p_3)^\nu
\nonumber\\
&+& \gamma_{a'a\,\mu} \gamma_{bb'\,\nu}(p_2-p_1)^\nu(p_4-p_3)^\mu
\biggr]v_{b'}(p_3)u_a(p_2)\,. 
\end{eqnarray}
which can be compared with the analogous amplitude given in
literature for the graviton case \cite{HLZ,Hewett}. On the other
hand, the explicit form of the cross section is given in  Appendix \ref{cs},
which agrees with \cite{Hewett,GRW}. These interactions would lead to 
modifications of the scattering cross sections such as
\begin{eqnarray}
e^+e^-  &\to& \ell \bar \ell,\ q \bar q\\
q\bar q &\to& \ell \bar \ell,\ q \bar q. \label{scatter}
\end{eqnarray}
where $\ell=e,\mu,\tau$. The Bhaba scattering, fermion
pair production in $e^+e^-$ colliders and Drell-Yan production at
hadron colliders have been studied in detail from the point of
view of the KK- graviton virtual exchange \cite{Hewett}. Dilepton
and dijet channels have been studied at LEP and dielectron
production at Tevatron. Also these processes are interesting for
$e^\pm p \to e^\pm p$ interactions observed at HERA .


For $\bar\psi(p_1),\psi(p_2)\longrightarrow
A^a_\mu(p_3),A^b_\nu(p_4)$ interactions (see Appendix B.2.), 
diphoton production have been studied at LEP
and Tevatron, whereas $WW$ and $ZZ$ production have been studied also 
at LEP. Exchange of virtual branons can also contribute to
processes like
\begin{eqnarray}
e^+e^-,\ q\bar q  &\to& \gamma\gamma,\ W^+W^-,\  ZZ\ {\rm and}\ gg \\
\gamma\gamma,\ gg &\to& \ell \bar \ell,\ q \bar q. \label{vv}
\end{eqnarray}
The contribution to the gauge boson production is given by
\begin{eqnarray}
{\cal M}^{2\psi\rightarrow2A}=\bar{v}_{a'}(p_1)V^{2\psi
2A}_{a'a\,\mu\nu}(-p_1, -p_2, p_3, p_4) u_a(p_2)
\epsilon^{\ast\,\mu}(p_3,\sigma_{b'})
\epsilon^{\ast\,\nu}(p_4,\sigma_b)\,. \label{m2f2a}
\end{eqnarray}
and to  fermion-antifermion production:
\begin{eqnarray}
{\cal M}^{2A\rightarrow2\psi}=\bar{u}_{a'}(p_1)V^{2\psi
2A}_{a'a\,\mu\nu}(p_1, p_2, -p_3, -p_4) v_a(p_2)
\epsilon^{\mu}(p_3,\sigma_{b'}) \epsilon^{\nu}(p_4,\sigma_b)\,.
\label{m2a2f}
\end{eqnarray}

In fact, for the graviton case, the
$e^+e^-\rightarrow\gamma\gamma,W^+W^-,ZZ$ processes have been
studied in  detail in \cite{AgDe}, as well as the $gg\rightarrow
l^+l^-$\cite{Hewett}. Moreover, explicit expresions for the
fermion-antifermion and diphoton production cross sections are
given in  Appendix \ref{cs}, or alternatively in \cite{GRW}.

Using the processes described above, we have obtained limits on
the parameter combination $f^2/(N^{1/4}\Lambda)$ from different
experiments at LEP, HERA and
Tevatron, which are
summarized in Table 2.
Also, with the same analogy with the Kaluza-Klein gravitons, 
 we can estimate the constraints from future
colliders. With that purpose, we have taken into account the
estimations calculated by Hewett \cite{Hewett} for future linear
colliders, Tevatron and LHC (see Table \ref{Future}).

\begin{table}[bt]
\centering
\begin{tabular}{||c|c c c||} \hline
\hline
Experiment          & $\sqrt s$ (TeV) & ${\cal L}$ (pb$^{-1}$) & $f^2/(N^{1/4}\Lambda)$ (GeV) \\ \hline
Tevatron-II$^{\,a,\,b}$  & 2.0        & $2 \times 10^3$        & 83                           \\ 
Tevatron-II$^{\,a,\,b}$  & 2.0        & $3 \times 10^4$        & 108                           \\ 
ILC$^{\,b}$         & 0.5             & $5\times 10^5$         & 261                          \\
ILC$^{\,b}$         & 1.0             & $2\times 10^5$         & 421                          \\ 
LHC$^{\,b}$     & 14              & $1\times 10^4$                 & 332                          \\
LHC$^{\,b}$     & 14              & $1\times 10^5$                 & 383                          \\ \hline
\hline
\end{tabular}
\caption{Estimated sensitivities on the parameter
$f^2/(N^{1/4}\Lambda)$  (results in GeV 
at the $95\;\%$ c.l.) for various colliders with
center-of-mass energies and integrated luminosities as indicated.
The indices $^{a,b,c}$ denote the same channels as in Table \ref{constrains1}.}
\label{Future}
\end{table}

\section{Two loops effects: Electroweak precision 
observables and muon anomalous magnetic moment}

Electroweak precision measurements are very useful to find
constraints on new physics models. The so called oblique
corrections (those corresponding to the $W$, $Z$ and $\gamma
$ two point functions) use to be described in terms of the $S,T,U$
\cite{STU} or the $\epsilon_1,\epsilon_2$ and $\epsilon_3$
parameters \cite{Alta}. The first order correction coming from the
Kaluza-Klein gravitons in the ADD models for rigid branes to the
parameter: $
\bar\epsilon\equiv \delta M_W^2/M_W^2 - \delta M_Z^2/M_Z^2 $
was computed in \cite{CPRS}. This result can be written as:
\begin{equation}
  \delta \bar\epsilon \simeq
  \frac{20\Lambda^2(M_Z^2-M_W^2)}{3(4\pi)^2}
  (5\,W_1+2\,W_2).
\end{equation}
Translating this result to our
context as in the previous section we find:
\begin{equation}
  \delta \bar\epsilon \simeq
   \frac{5\,(M_Z^2-M_W^2)}{12\,(4\pi)^4}
  \frac{ N\Lambda^6}{f^8}
\end{equation}

Notice that this is in fact a two-loop result since it is obtained
from a one-loop computation by using an effective
Lagrangian which is coming from another one-loop computation.

The experimental value of $\bar\epsilon$ obtained from LEP \cite{EWWG}
is $\bar\epsilon=(1.27 \pm 0.16)\times 10^{-2}$. This value is consistent with
the SM prediction for a light higgs $m_H\leq 237$ GeV at 95 \% c.l.
On the other hand, the theoretical uncertainties are one order of magnitud smaller
\cite{Alta} and therefore, we can estimate the constraints for the branon contribution
at 95 \% c.l. as $|\delta\bar\epsilon|\leq 3.2\times 10^{-3}$.
Thus it is possible to set the bound:
\begin{equation}\label{eb}
    \frac{f^4}{ N^{1/2}\Lambda^3}\geq \;3.1\; \mbox{GeV (95 \% c.l.)}
\end{equation}
This result has a
stronger dependence on $\Lambda$ ($\Lambda^6$) than the
interference cross section between the branon and SM interactions
($\Lambda^4$). Therefore, the constraints coming from this analysis are
complementary to the previous ones.

A further constraint to the branon parameters can be obtained
from the  $\mu$ anomalous magnetic moment.  The first branon contribution
to this parameter can be obtained  from a one loop computation with the 
Lagrangian
 given by (\ref{eff}). The result for the KK graviton tower was first calculated  by
\cite{Graesser} and confirmed by \cite{CPRS} in a different way, 
and can be written as:

\begin{equation}
\delta a_\mu \simeq \frac{2 m_\mu^2 \Lambda^2}{3(4\pi)^2}
  (11\,W_1-12\,W_2),
\end{equation}
which for the branon case can be translated into:
\begin{equation}\label{gb}
\delta a_\mu \simeq \frac{5\, m_\mu^2}{114\,(4\pi)^4}
  \frac{N\Lambda^6}{f^8}.
\end{equation}
This expression is is qualitatively similar to other 
$g-2$ contributions obtained in
different analyses of extra-dimension models
\cite{branemuon}.
The result depends on the cut-off $\Lambda$ in the same way as 
the electroweak precision parameters. However the experimental
situation is slightly different. In a sequence of increasingly
more precise measurements, the  821 Collaboration at the
Brookhaven Alternating Gradient Syncrotron has reached a fabulous
 relative precision of 0.5 parts per million in the determination
 of $a_\mu=(g_\mu-2)/2$ \cite{BNL}. These measurements provide a
 stringent test not only of new physics but also of the SM. Indeed,
 the present result is only marginally consistent with the SM. Taking
 into account the $e^+e^-$ collisions to calculate the $\pi^+\pi^-$
 spectral functions, the deviation with respect to the SM prediction is at
 $2.6$ standard deviations \cite{gm2}. In particular: $\delta a_\mu \equiv a_\mu
(exp) - a_\mu (SM) =(23.4 \pm 9.1)\times 10^{-10}$.
Using Equation (\ref{gb}) we can estimate the {\it preferred} parameter region for branons
 to provide the observed difference:
\begin{equation}\label{mb}
    6.0\; \mbox{GeV}\;\geq\,\frac{f^4}{ N^{1/2}\Lambda^3}\,
\geq\;\mbox{2.2  GeV (95 \% c.l.)}
\end{equation}
We observe that the correction to the muon anomalous
magnetic moment is in the right direction and that
 it is possible to avoid the present constraints and
 improve the observed experimental value obtained by the E821 Collaboration.

 If there
 is  really new physics in the muon anomalous
magnetic moment,  and this new physics is due to branon radiative
 corrections, the phenomenology
of these particles should be observed at the LHC
and in a possible future ILC (see Table \ref{Future}). In particular,
the LHC should observe an important difference in the 
channels: $pp\rightarrow e^+e^-$ and
 $pp\rightarrow \gamma \gamma$
with respect to the SM prediction. The ILC should observe 
the most important effect in
 the process:
$e^+e^-\rightarrow e^+e^-$. Moreover, in \cite{CDMmuon} it was shown
that the same parameter region in which branons could explain the 
muon anomalous magnetic moment is also  compatible with
a cosmological branon relic abundance enough to account for the  
observed dark matter relative density  \cite{WMAP}.

\section{Conclusions}
In this work we have studied the phenomenological consequences of branon radiative
corrections, by calculating the one-loop effective action for SM particles,
obtained after
integrating the branon fields out. We have found new interaction vertices
of SM particles, in particular new four-fermion interactions and interactions
involving two fermions and two gauge fields. Our results, computed with a
cutoff regulator, are very similar to those
obtained by integrating  the graviton KK modes at the tree level in ADD models,
and, accordingly,  we have translated the different constraints to 
the branon case.
Thus,
we have obtained limits for the combination of parameters 
$f^2/(N^{1/4}\Lambda)$
from present experiments at LEP and Tevatron, and also for future colliders
(ILC and LHC).

We have also considered the branon two-loop effect on electroweak precision observables
and on the muon anomalous magnetic moment. We have evaluated the corresponding corrections and
obtained the preferred parameter range for branons in order to fit the Brookhaven
results, and at the same time, to be consistent with LEP precision measurements.
In Fig. 1 we include all those limits and also the parameter
region in which the theory can be considered as strongly interacting, i.e.
 ($\Lambda\,\gsim\,4\sqrt{\pi}fN^{-1/4}$) and  for which
the loop expansion is no longer valid.

\begin{figure}[ht]
\begin{center}
\epsfxsize=10cm   
\epsfbox{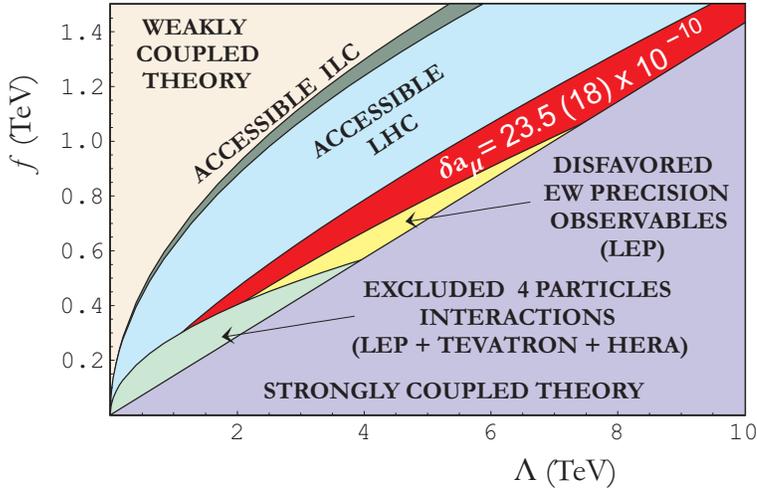}
\caption {\footnotesize Main limits from branon radiative
corrections in the $f-\Lambda$ plane for a model with $N=1$.
The (red) central area shows the region in which the branons account for
the
muon magnetic moment deficit observed by the E821 Collaboration
\cite{BNL,gm2}, and
at the same time, are consistent with present collider experiments
(whose main constraint
comes from the Bhabha scattering at LEP) and  electroweak precision
observables. Prospects for future colliders are also plotted.}
\label{CDM}
\end{center}
\end{figure}

\newpage
\appendix
\section{Divergent integrals}
\label{Integrals}
Definitions of various divergent integrals used in the text:
\begin{eqnarray}
J^{(0)}(p)&=&\int d\tilde q
\frac{1}{(q^2-M^2+i\epsilon)((p+q)^2-M^2+i\epsilon)}
.
\end{eqnarray}
\begin{eqnarray}
J^{(2)}_{\mu\nu}(p)&=&\int d\tilde q \frac{q_\mu
q_\nu}{(q^2-M^2+i\epsilon)((p+q)^2-M^2+i\epsilon)}
.
\end{eqnarray}
\begin{eqnarray}
J^{(M)}_{\mu\nu\rho\sigma}(p)&=&\int d\tilde q \frac{q_\mu q_\nu
(p+q)_\rho (p+q)_\sigma}{(q^2-M^2+i\epsilon)((p+q)^2-M^2+i\epsilon)}
.
\end{eqnarray}

\section{Effective Feynman rules}
\label{Vertex}

In this section we give the most important effective
Feynman rules, obtained by the integration of the
branons. We are going to present the
fundamental new vertices with outgoing momenta from (\ref{eff}).

\subsection{Effective 4-fermion vertex}


One of the most relevant contribution of virtual branons to the
phenomenology of the SM particles is the effect on
four-fermion interactions. For a generic four-fermion process, the
branons induce a new effective vertex of the form:
%
%
%


\begin{figure}[h]
\begin{center}
\epsfxsize=13cm   
\epsfbox{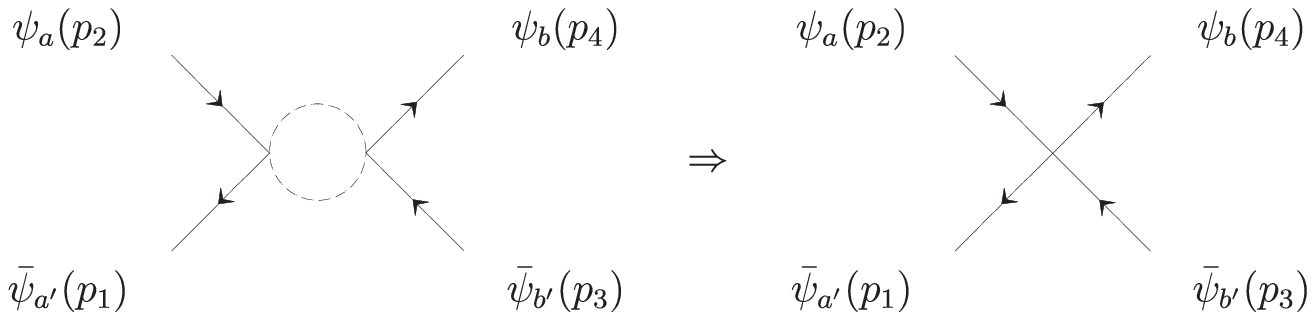}
\label{radplot1}
\end{center}
\end{figure}

\begin{eqnarray}
V^{4\psi D}_{a'abb'}(p_1, p_2, p_3, p_4)
&=&\frac{W_1}{4}
\biggl[32m_{\psi_a}\delta_{a'a}m_{\psi_b}\delta_{bb'}
+\gamma_{a'a\,\mu}\gamma_{bb'}^\mu(p_2-p_1)_\nu(p_4-p_3)^\nu\nonumber\\
&+&12 m_{\psi_b}\delta_{bb'}\gamma_{a'a\,\mu}(p_2-p_1)^\mu+12
m_{\psi_a}\delta_{a'a}\gamma_{bb'\,\mu}(p_4-p_3)^\mu\nonumber\\
&+&(\gamma_{a'a\,\mu} \gamma_{bb'\,\nu}+4\gamma_{a'a\,\nu}
\gamma_{bb'\,\mu})(p_2-p_1)^\nu(p_4-p_3)^\mu\biggr]\nonumber\\
&+&\frac{W_2}{2}\,[8m_{\psi_a}\delta_{a'a}+3\gamma_{a'a\,\mu}(p_2-p_1)^\mu]
\nonumber\\
&&[8m_{\psi_b}\delta_{bb'}+3\gamma_{bb'\,\nu}(p_4-p_3)^\nu]\,.
%
\end{eqnarray}

In the case in which the fermion fields $a$ and $b$ are the same, one
has to take into account two different effects in order to obtain the vertex
from the above expression. On one hand, a factor of
$2$  due to the quadratic term in the SM energy-momentum tensor is not present,
and on the other hand, the symmetrization with respect to the
change: $\{p_1,p_2\}\leftrightarrow\{p_3, p_4\}$ should be
performed. Therefore, the general form of the
four-fermion vertex is given by:

\begin{eqnarray}
V^{4\psi}_{a'abb'}(p_1, p_2, p_3, p_4)
&=&V^{4\psi D}_{a'abb'}(p_1, p_2, p_3, p_4)\\
&+&{1\over 2 }\delta_{ab}\delta_{a'b'}
\biggl[V^{4\psi D}_{bb'a'a}(p_3, p_4,p_1,
p_2)-V^{4\psi D}_{a'abb'}(p_1, p_2,p_3, p_4)\biggl]\,.
\nonumber
\end{eqnarray}

\subsection{Effective $\bar\psi,\psi,A_\mu,A_\nu$ vertex}
The exchange of virtual branons can also contribute to processes
involving both a fermion pair and a gauge field pair
As shown in the diagram,  branons induce a new effective
vertex, which, with the same momenta assignment as before, takes the form:
%
%
%


\begin{figure}[h]
\begin{center}
\epsfxsize=13cm   
\epsfbox{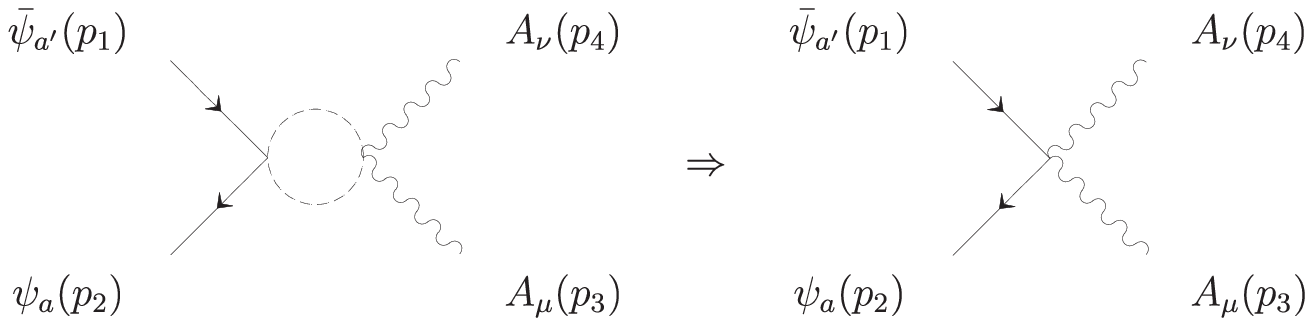}
\label{radplot2}
\end{center}
\end{figure}

\begin{eqnarray}
V^{2\psi 2A}_{a'a\,\mu\nu}(p_1, p_2, p_3, p_4)
&=&W_1\biggl[
[(p_2-p_1)^\lambda(p_{3\lambda}p_{4\mu})
-(p_4,p_3)(p_2-p_1)_\mu]\gamma_{a'a\,\,\nu}
\nonumber\\
&+&[(p_2-p_1)^\lambda(p_{4\lambda}p_{3\nu})
-(p_4,p_3)(p_2-p_1)_\nu]\gamma_{a'a\,\,\mu}
\nonumber\\
&-&[(p_2-p_1)^\lambda\gamma_{a'a}^\sigma
(p_{4\lambda}p_{3\sigma}-p_{3\lambda}p_{4\sigma}-\eta_{\lambda\sigma}(p_4,p_3))]
\eta_{\mu\nu}\nonumber\\
&-&[(p_2-p_1)_\lambda p_{4\mu}p_{3\nu} -(p_2-p_1)_\mu
p_{4\lambda}p_{3\nu}\nonumber\\
&&-(p_2-p_1)_\nu
p_{3\lambda}p_{4\mu}]\gamma_{a'a}^\lambda\\
&-& m_A^2[(4m_\psi\delta_{a'a}+\gamma_{a'a}^\lambda(p_2-p_1)_\lambda)\eta_{\mu\nu}
\nonumber\\
&&+(p_2-p_1)_\mu\gamma_{a'a\,\,\nu}+(p_2-p_1)_\nu\gamma_{a'a\,\,\mu}]
\biggr]\nonumber\\
&-&2\,W_2\,m_A^2[8m_\psi\delta_{a'a}+3\gamma_{a'a}^\lambda(p_2-p_1)_\lambda]\eta_{\mu\nu}
\,. \label{V2f2A}\nonumber
\end{eqnarray}

\subsection{Effective $B_\rho,B_\sigma,A_\mu,A_\nu$ vertex}
Another effective interaction produced by virtual branon exchange
can also contribute to processes involving two different gauge
field pairs.
Indeed,  branons induce the following effective vertex:
%
%
%


\begin{figure}[h]
\begin{center}
\epsfxsize=13cm   
\epsfbox{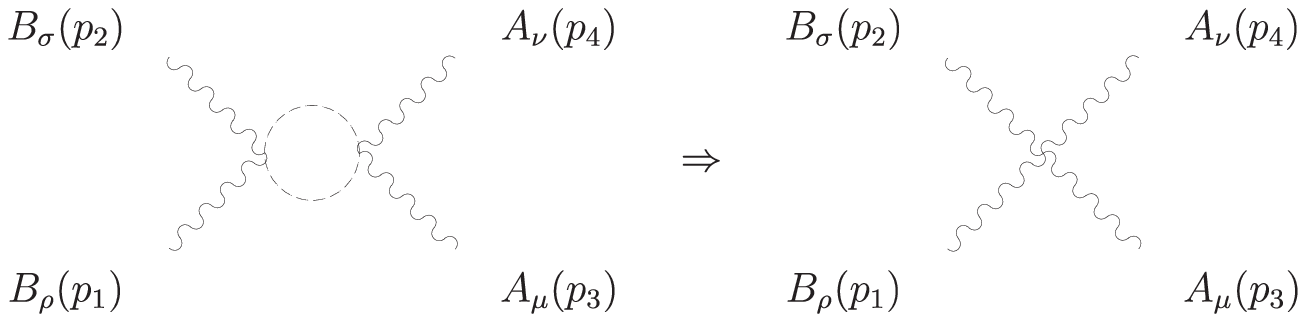}
\label{radplot3}
\end{center}
\end{figure}

\begin{eqnarray}
V^{2B2A}_{\mu\nu\rho\sigma}(p_1, p_2, p_3, p_4)
&=&4\,W_1\biggl[
p_{2\mu}p_{3\sigma}(p_{1\rho}p_{4\nu}-p_{1\nu}p_{4\rho})
\nonumber\\
&+&p_{1\mu}p_{3\rho}(p_{2\sigma}p_{4\nu}-p_{2\nu}p_{4\sigma})
+p_{3\mu}(p_{2\sigma}p_{4\rho}p_{1\nu}-p_{2\nu}p_{4\sigma}p_{1\rho})
\nonumber\\
&-&(p_3,p_4)[p_{2\mu}(-p_{1\nu}\eta_{\sigma\rho}+p_{1\rho}\eta_{\nu\sigma})+
p_{1\mu}(-p_{2\nu}\eta_{\sigma\rho}\nonumber\\
&+&
p_{2\rho}\eta_{\nu\sigma})+p_{2\nu}p_{1\rho}\eta_{\sigma\mu}+p_{1\nu}p_{2\sigma}\eta_{\mu\rho})
-2p_{1\rho}p_{2\sigma}\eta_{\mu\nu}]
\nonumber\\
&-&(p_1,p_2)[p_{4\nu}(p_{3\sigma}\eta_{\mu\rho}+p_{3\rho}\eta_{\mu\sigma})+
p_{3\mu}(p_{4\rho}\eta_{\sigma\nu}+p_{4\sigma}\eta_{\nu\rho})\nonumber\\
&-& 2p_{3\mu}p_{4\nu}\eta_{\sigma\rho}
-(p_{4\rho}p_{3\sigma}+p_{3\rho}p_{4\sigma})\eta_{\mu\nu}]
\nonumber\\
&-&(p_2,p_4)[p_{3\sigma}(p_{1\rho}\eta_{\mu\nu}-p_{1\nu}\eta_{\mu\rho})+
p_{3\mu}p_{1\nu}\eta_{\rho\sigma}]\nonumber\\
&-&(p_2,p_3)[p_{4\sigma}(p_{1\rho}\eta_{\mu\nu}-p_{1\mu}\eta_{\nu\rho})+
p_{4\nu}p_{1\mu}\eta_{\rho\sigma}]
\nonumber\\
&-&(p_1,p_4)[p_{3\rho}(p_{2\sigma}\eta_{\mu\nu}-p_{2\nu}\eta_{\mu\sigma})+
p_{3\mu}p_{2\nu}\eta_{\rho\sigma}]
\nonumber\\
&-&(p_1,p_3)[p_{2\mu}
(p_{4\nu}\eta_{\rho\sigma}-p_{4\rho}\eta_{\nu\sigma})+
p_{4\rho}p_{2\sigma}\eta_{\mu\nu}]
\nonumber\\
&-&(p_1,p_3)[p_{2\mu}
(p_{4\nu}\eta_{\rho\sigma}-p_{4\rho}\eta_{\nu\sigma})+
p_{4\rho}p_{2\sigma}\eta_{\mu\nu}]
\nonumber\\
&+&(p_1,p_4)(p_2,p_3)
[\eta_{\rho\nu}\eta_{\mu\sigma}+\eta_{\rho\mu}\eta_{\nu\sigma}
-2\eta_{\rho\sigma}\eta_{\mu\nu}]\nonumber\\
&+&
\eta_{\rho\sigma}\eta_{\mu\nu}[(p_1,p_4)(p_2,p_3)+(p_1,p_3)(p_2,p_4)]
\nonumber\\
&+& m_A^2[-p_{4\mu}(-2p_{3\nu}\eta_{\rho\sigma}
+p_{3\rho}\eta_{\nu\sigma}+p_{3\sigma}\eta_{\nu\rho})\\
&-&p_{3\nu}(p_{4\rho}\eta_{\mu\sigma}+p_{4\sigma}\eta_{\mu\rho})
+{1\over 2}
\eta_{\mu\nu}(p_{4\rho}p_{3\sigma}+p_{3\rho}p_{4\sigma})\nonumber\\
&+&(p_3,p_4)(\eta_{\nu\sigma}\eta_{\mu\rho}
+\eta_{\mu\sigma}\eta_{\rho\nu}-\eta_{\rho\sigma}\eta_{\mu\nu})]
\nonumber\\
&+& m_B^2[-p_{2\rho}(-2p_{1\sigma}\eta_{\mu\nu}
+p_{1\mu}\eta_{\sigma\nu}+p_{1\nu}\eta_{\sigma\mu})\nonumber\\
&-&p_{1\sigma}(p_{2\mu}\eta_{\rho\nu}+p_{2\nu}\eta_{\mu\rho})
+{1\over 2} \eta_{\rho\sigma}(p_{2\mu}p_{1\nu}+p_{1\mu}p_{2\nu})\nonumber\\
&+&(p_1,p_2)(\eta_{\nu\sigma}\eta_{\mu\rho}
+\eta_{\mu\sigma}\eta_{\rho\nu}-\eta_{\rho\sigma}\eta_{\mu\nu})]
\nonumber\\
&+& m_A^2 m_B^2[\eta_{\nu\sigma}\eta_{\mu\rho}
+\eta_{\mu\sigma}\eta_{\rho\nu}]
\biggr]+
%
2\,W_2\,m_A^2 m_B^2\eta_{\mu\nu}\eta_{\rho\sigma}
. \label{V2B2A}\nonumber
\end{eqnarray}

In the case in which  the gauge bosons $A$ and $B$ are the same, one
should again take into account the points commented above. In particular
the symmetrization with respect to the change:
$\{p_1,p_2,\mu,\nu\}\leftrightarrow\{p_3, p_4,\rho,\sigma\}$
should be performed. Therefore, the vertex with four identical
gauge boson can be written as:

\begin{eqnarray}
V^{4A}_{\mu\nu\rho\sigma}(p_1, p_2, p_3, p_4)
={1\over 2}[V^{2B2A}_{\mu\nu\rho\sigma}(p_1, p_2, p_3, p_4)
+V^{2B2A}_{\rho\sigma\mu\nu}(p_3, p_4,p_1, p_2)]
.
\end{eqnarray}

\section{Cross sections}
\label{cs}

In this section we show the modifications in the 
cross-section of four-particles processes derived from 
virtual branon exchange in terms of 
$s\equiv (p_1+p_2)^2$, $t\equiv (p_1-p_3)^2$ and $u\equiv (p_2-p_3)^2$. 
We are neglecting the masses of these
particles, which means $s+t+u=0$.

\subsection{
$\sigma_1: f(p_1)\bar f(p_2) \to \gamma(p_3)\gamma(p_4)$}

The diphoton production by 
fermion-antifermion annihilation with electric charge 
$Q_f$ and number of colours $N_f$ is given by:

\begin{eqnarray}
\frac{d\sigma_1}{dt}=\frac{s^2-2tu}{N_f s^2}
\left[\frac{2\pi\alpha Q_f^2 }{\sqrt{tu}}
+\frac{W_1}{2}\sqrt{tu}\right]^2\,.
\end{eqnarray}

\subsection
{$\sigma_2: g(p_1)g(p_2) \to l^+(p_3)l^-(p_4)$}

On the contrary, the dilepton production by gluon annihilation 
does not present interference term:

\begin{eqnarray}
\frac{d\sigma_2}{dt}=\frac{W_1^2\,tu}{64\pi s^2}\left[s^2-2tu\right]\,.
\end{eqnarray}

\subsection{
$\sigma_3: g(p_1)g(p_2)\to \gamma(p_3)\gamma(p_4)$}


The situation is similar for the diphoton production 
by gluon annihilation, since there is no SM contribution at tree-level:

\begin{eqnarray}
\frac{d\sigma_3}{dt}=\frac{W_1^2}{64\pi s^2}
\left[s^4-2tu(2s^2-tu)\right]\,.
\end{eqnarray}

\subsection{
$\sigma_4: e^-(p_1) e^+(p_2) \to f(p_3)\bar f(p_4)\;\;(f\neq \nu_e, e^-)$}

To illustrate the four-fermion interaction contribution, 
we can write the cross-section for the fermion-antifermion 
production (except $\nu_e$ and $e^-$) 
in $e^+ e^-$ collisions in terms of the vector 
$v_f=T_f-2Q_f\sin^2\theta_W$ and axial 
$a_f=T_f$ couplings of the particular fermion field:  

\begin{eqnarray}
\frac{d\sigma_4}{dt}&=&\left.\frac{d\sigma_4}{dt}\right\vert_{\rm SM}
+\frac{N_f W_1^2}{128\pi s^2}\left[s^4-2tu(5s^2-16tu)\right]
\nonumber\\
 &-&\frac{N_f\alpha W_1}{4s^3}\biggl\{ Q_e Q_{f}(t-u)^{3}
+\frac{1}{\sin^2 2\theta_W}
\\&&
\frac{s}{s-M_Z^2}
\left[ v_ev_{f}(t-u)^{3}+a_ea_{f}s(s^2-6tu)\right] \biggr\}\,.\nonumber 
\end{eqnarray}

\subsection{
$\sigma_5: e^-(p_1) e^+(p_2) \to e^-(p_3) e^+(p_4)$}

For the Bhabha scattering, the cross-section presents 
more terms since one has to take into account the $t$-channel contributions:

\begin{eqnarray}
\frac{d\sigma_5}{dt}&=&\frac{d\sigma_4}{dt}
+\frac{W_1^2}{128\pi s^2}\left[40s^4+6t(31st^2-21s^3-40s^2u)+9t^4\right]
\nonumber\\
&-&\frac{\alpha W_1}{4s^3}\biggl\{
\frac{Q_e^2}{t} \left[9s^4+22ts^3+24t^2s^2-11t^3u-10t^4\right]
\\
&+&
\frac{s}{\sin^2 2\theta_W}\frac{v^2_e+a^2_e}{s-M_Z^2}\left[u(4t^2-4s^2+5tu)\right]
+\frac{1}{\sin^2 2\theta_W}\frac{s}{t-M_Z^2}
\nonumber \\
&& \left[ v^2_e(8s^3+6ts^2)+(v^2_e+a^2_e)[s^3+12s^2t-5t^2(3u+2t)]\right]\biggr\}\,.
\nonumber
\end{eqnarray}


All these results are in agreement with the expressions calculated 
for other kind of models which predict the same Lagrangian (\ref{eff}). 
In particular, for  KK-gravitons in the ADD model: 
$W_2= -W_1/(N+2)$ and $W_1$ is related to a new energy scale: $M_S$
\cite{Hewett,GRW,AgDe}. For example, $W_1=
4\lambda/M_S^4$ in \cite{Hewett}
(where typically $\lambda=\pm 1$,
takes into account the unknownness of the exact theory). 
So we can use the effective vertices and cross-sections given 
in the Apendices
\ref{Vertex} and \ref{cs} for branons or gravitons using the 
corresponding
definitions of the parameters $W_1$ and $W_2$. In fact, we can
estimate directly the bounds over $f^2/(\Lambda N^{1/4})$ using
the bounds over $M_S$.  In the most interesting cases, 
the contribution of the term proportional to
$W_2$ is zero or neglegible and it is a good
estimation to take, for $\lambda=1$:
\begin{eqnarray}
\frac{f^2}{\Lambda N^{1/4}}=\frac{M_S}{4(24\pi^2)^{1/4}}\simeq
0.064\, M_S.
\end{eqnarray}
Typically, when $f\ll M_D$ the most important signal of brane worlds comes
from branons. In such a case, we can estimate $f^2/(\Lambda
N^{1/4})$ as it is shown in Table \ref{constrains1}.

{\em Acknowledgments} ---
J. A. R. C. thanks C. P. Martin for important comments on the
 renormalization procedure. This work is supported in part
by DGICYT (Spain) under project numbers
FPA 2004-02602 and FPA 2005-02327, by NSF grant No.~PHY--0239817
and by the Fulbright-MEC (Spain) program.

\end{document}